\documentclass[preprint,amsmath, amssymb]{revtex4}
\usepackage{fontenc}
\usepackage[latin1]{inputenc}
\usepackage{graphicx}

\begin{document}

\preprint{FAU-TP3-08-06}

\title{Quantization of Sine-Gordon solitons on the circle: semiclassical vs. exact results}
\author{Michael Pawellek}
\affiliation{Institut für Theoretische Physik III, \\ Universität Erlangen-Nürnberg, \\ Staudtstr.7, D-91058 Erlangen, Germany}
\email{michi@theorie3.physik.uni-erlangen.de}

\begin{abstract}
We consider the semiclassical quantization of sine-Gordon solitons on the circle with periodic and anti-periodic boundary conditions.
The 1-loop quantum corrections to the mass of the solitons are determined using zeta function regularization in the integral representation.
We compare the semiclassical results with exact numerical calculations in the literature and find excellent agreement even outside the plain 
semiclassical regime.
\end{abstract}
\maketitle

\section{Introduction}

The semiclassical quantization method is a fruitful technique to explore non-perturbative
properties of quantum field theories \cite{Das2, Neve}. The determination of quantum corrections to the mass of the \(\phi^4\) kink
and the sine-Gordon soliton are classic textbook examples of this method \cite{Raja}.
Although the sine-Gordon model is integrable \cite{Zamo} and the \(\phi^4\) model is not \cite{Makh}, on the semiclassical level these theories are surprisingly similar in the
one soliton/kink sector. In both cases the fluctuation equation obtained after expansion around the classical soliton/kink solutions are exactly solvable 
reflectionless Schroedinger equations of the P\"oschl-Teller type \cite{Raja}. 
Considering these models on a compact space, e.g. a circle, one gets instead \textit{quasi-exactly} 
solvable \cite{Turb} finite gap Schroedinger equations of the Lam\'e type \cite{Mus3, Bra1}. In this cases only a finite number of (anti-)periodic 
eigenfunctions  and -values can be determined analytically \cite{Arsc}.

For quantum corrections to the energy of non-trivial field configurations one needs at first sight information of the full fluctuation spectrum.
In \cite{Pawe} the special finite-gap properties of the \(n=2\) Lam\'e equation \cite{Arsc} and the integral representation of the spectral
zeta function \cite{Kir1, Kir3, Kir4} were used to construct an analytic
result for the 1-loop quantum corrections to the mass of the twisted \(\phi^4\) kink on the circle without explicit knowledge
of the spectrum. It was found that an energetically preferred radius exists, where the contributions of the classical and
1-loop part are of the same order of magnitude. Therefore the question arises if higher loop corrections may spoil this picture.

To settle this question we will not consider higher loop effects directly, but take a different route. We will consider the sine-Gordon model on \(S^1\), 
since it is very similar in the semiclassical approximation to the \(\phi^4\) theory. The fluctuation equation of (anti-)periodic solitons on \(S^1\) of the sine-Gordon model is the \(n=1\) Lam\'e equation \cite{Mus3}. Therefore we can 
apply the techniques used for the \(\phi^4\) model \cite{Pawe} also in this case. The integrability enables us to compare semiclassical results with 
exact results of the soliton energy, obtained in \cite{Fev2, Feve} by numerically solving the corresponding non-linear-integral-equations 
(NLIE) \cite{Dest}. This will give new insights into the question on relevance of higher loop corrections on \(S^1\) for the sine-Gordon soliton.

We will concentrate in the following on the one soliton sector on the compact manifold \(S^1\), where we can impose two different boundary conditions:
\begin{itemize}
 \item periodic b.c.: \(\phi(x+R)=\phi(x)+\frac{2\pi}{\beta}\)
 \item anti-periodic b.c.: \(\phi(x+R)=-\phi(x)+\frac{2\pi}{\beta}\)
\end{itemize}
In the past only asymptotic expressions of the semiclassical 1-loop energy for \(k\to 0\) and \(k\to 1\) of the elliptic modulus were obtained 
\cite{Bra1, Mus3, Muss}. We will give analytic results valid for all \(k\) and therefore \(R\). 

First we review the classical solutions \cite{Mus3} of the corresponding b.c. and then use the spectral discriminant of the \(n=1\) Lam\'e equation 
\cite{Bra1} to determine the 1-loop contributions. Finally we compare our result with numerical calculations, 
which used the integrability of sine-Gordon model \cite{Feve, Fev2}.

\section{Classical solutions}

We consider the sine-Gordon model with Lagrangian
\begin{equation}\label{eq:Lagrange}
 \mathcal{L}=\frac{1}{2}\partial_{\mu}\phi\partial^{\mu}\phi- V(\phi),\qquad V(\phi)=\frac{m^2}{\beta^2}(1-\cos(\beta\phi)).
\end{equation}
with spatial direction compactified on a circle with circumference \(R\).
We review the properties of classical static solutions \cite{Muss,Mus3} of the equation of motion following from (\ref{eq:Lagrange}) on \(S^1\).

\subsection{Periodic boundary condition}
With (quasi-) periodic boundary conditions \(\phi(x+R)=\phi(x)+\frac{2\pi}{\beta}\) the static field configuration is given by
\begin{equation}\label{eq:Sinegordonperi}
 \phi_0(x)=\frac{\pi}{\beta}+\frac{2}{\beta}\mathrm{am}\left(\frac{m(x-x_0)}{k},k\right),
\end{equation}
where the solution depends implicitly on the radius by
\begin{equation}
 R=\frac{2k}{m}\mathbf{K}(k).
\end{equation}
The classical energy can be expressed in terms of complete elliptic integrals of first and second kind:
\begin{equation}\label{eq:Energy1}
 E_{cl}(k)=\frac{4m}{\beta^2k}\left[(k^2-1)\mathbf{K}(k)+2\mathbf{E}(k)\right].
\end{equation}


For \(k\to 1\) the radius goes \(R\to\infty\) and the periodic solution (\ref{eq:Sinegordonperi}) reduces to the standard Sine-Gordon soliton
\begin{equation}
 \phi_0(x)\to\frac{4}{\beta}\arctan\left(e^{m(x-x_0)}\right)
\end{equation}
and the energy (\ref{eq:Energy1}) becomes the classical soliton mass in this limit:
\begin{equation}
 E_{cl}(k)\to\frac{8m}{\beta^2}.
\end{equation}

\subsection{Anti-periodic boundary conditions}
With (quasi-) anti-periodic boundary conditions \(\phi(x+R)=-\phi(x)+\frac{2\pi}{\beta}\) the solutions is given by
\begin{equation}\label{eq:Sinegordonaperi}
 \phi_0(x)=\frac{2}{\beta}\arccos\left(k\;\mathrm{sn}(m(x-x_0),k)\right),
\end{equation}
where the solution depends implicit on the radius 
\begin{equation}
 R=\frac{2}{m}\mathbf{K}(k).
\end{equation}
The classical energy can be expressed in terms of complete elliptic integrals of first and second kind:
\begin{equation}\label{eq:Energy2}
 E_{cl}(k)=\frac{4m}{\beta^2}\left[(k^2-1)\mathbf{K}(k)+2\mathbf{E}(k)\right].
\end{equation}


As in the case of the twisted \(\phi^4\) kink \cite{Pawe} a critical radius exists at
\begin{equation}
 R_0=\frac{\pi}{m},
\end{equation}
where the soliton solution (\ref{eq:Sinegordonaperi}) reduces to the constant field configuration
\begin{equation}\label{eq:constfield}
 \phi_0(x)\to\frac{\pi}{\beta},\qquad k\to 0,
\end{equation}
and the the energy (\ref{eq:Energy2}) becomes in this limit
\begin{equation}
 E_{cl}\to\frac{2\pi m}{\beta^2},
\end{equation}
which coincides with the energy 
\begin{equation}
 E_{cl}(R)=\frac{2m^2}{\beta^2}R
\end{equation}
of the constant field configuration (\ref{eq:constfield}) at \(R=R_0\).

Again, for \(k\to 1\) the radius \(R\to\infty\) and the anti-periodic solution (\ref{eq:Sinegordonaperi}) reduces to the standard 
Sine-Gordon soliton:
\begin{equation}
 \phi_0(x)\to\frac{4}{\beta}\arctan\left(e^{m(x-x_0)}\right)
\end{equation}
and the energy (\ref{eq:Energy2}) becomes the classical soliton mass in this limit:
\begin{equation}
 E_{cl}(k)\to\frac{8m}{\beta^2}.
\end{equation}

\section{1-loop Contributions}
Expanding in the Lagrangian (\ref{eq:Lagrange}) the field \(\phi(x)=\phi_0(x)+e^{i\sqrt{\lambda}t}\chi(x)\) about a certain classical field configuration 
\(\phi_0(x)\) leads to a corresponding fluctuation equation. All energies of the sine-Gordon soliton states are measured relative to the Minkowski 
vacuum \(\phi_0=0\) without nontrivial boundary conditions. The fluctuation equation reads in this case:
\begin{equation}\label{eq:ZetaResolvent}
 \left[-\frac{\mathrm{d}^2}{\mathrm{d}x^2}+m^2\right]\chi(x)=\lambda\chi(x),
\end{equation}
or 
\begin{equation}
 -\frac{\mathrm{d}^2}{\mathrm{d}x^2}\chi(x)=\kappa^2\chi(x),
\end{equation}
when introducing the momentum-like parameter 
\begin{equation}\label{eq:momentum}
 \kappa^2=\lambda-m^2.
\end{equation}
The mass of the elementary quanta in this vacuum are \(m\).

\subsection{Spectral zeta functions}
In order to fix the notation we give in this section a short summary of zeta function regularization and the integral representation of spectral 
zeta functions \cite{Kir1, Kir3, Kir4}.

For the eigenvalue problem 
\begin{equation}\label{eq:DiffEq}
 D\phi(x,\lambda)=\lambda\phi(x,\lambda)
\end{equation}
with a second order differential operator \(D=-\partial_x^2+V(x)\) and properly chosen boundary conditions, the set of eigenvalues \(\{\lambda_i\}_{i\in\mathbf{N}}\)
is discrete and bounded from below. If (\ref{eq:DiffEq}) is a fluctuation equation obtained by a semiclassical expansion the 1-loop energy contribution 
to the classical solution is given by
\begin{equation}
 E_{1-loop}=\frac{1}{2}\sum_{n=0}^{\infty}\sqrt{\lambda_n}.
\end{equation}
In quantum field theories this expression is divergent and has to be regularized. In zeta function regularization one works with the spectral zeta 
function formally defined by
\begin{equation}\label{eq:Zetadef}
 \zeta_D(s)=\mu^{1+2s}\sum_{n=1}^{\infty}\lambda_n^{-s},
\end{equation}
with \(\mathrm{Re}(s)>s_0\), where \(s_0\) depends e.g. on the numbers of dimensions. The parameter \(\mu\) with dimension of mass is introduced in 
order that the energy has the correct dimension for all values of \(s\). The 1-loop contribution to the energy of
a classical field configuration in zeta function regularization is then defined as the value of the analytic continuation of 
\(\zeta_D(s)\) at \(s=-\frac{1}{2}\):
\begin{equation}
 E_{1-loop}=\frac{1}{2}\zeta_D(-1/2).
\end{equation}
For renormalization we will apply the large mass subtraction scheme, which is widely used in Casimir energy calculations \cite{Bor2}. For a physical 
field with mass \(m\) one expects that all quantum fluctuations will be suppressed in the limit of large mass \(m\), because for a field with infinite 
mass the quantum fluctuations should vanish. So one expects that for \(m\to\infty\) there are no 1-loop corrections at all and a good 
renormalization condition is \cite{Bor4, Bor2}
\begin{equation}\label{eq:RenormCond}
 E_{ren}\to 0 ,\;\;\text{for}\;\;m\to\infty.
\end{equation}
With this prescription at hand one can identify and subtract the divergent (when \(s=-\frac{1}{2}\) is a pole of
\(\zeta_D(s)\)) contributions \(E_{div}(s)\) from \(E_{1-loop}(s)\) and the renormalized 
energy is then given by
\begin{equation}
 E_{ren}=\lim_{s\to-\frac{1}{2}}\left[E_{1-loop}(s)-E_{div}(s)\right].
\end{equation}
 
Assume we have a function \(\Delta(\lambda)\), whose zeros of n-th order are at the positions \(\lambda_i>0\) of the n-fold degenerate 
eigenvalues of the spectral problem under consideration:
\begin{equation}
 \Delta(\lambda)=0\;\; \Leftrightarrow\;\;\; \lambda\;\;\text{eigenvalue of }\;\; D.
\end{equation}
Such a function is called the spectral discriminant. 
Then one can write the spectral zeta function as a contour integral
\begin{equation}\label{eq:ZetaInt1}
 \zeta_D(s)=\frac{1}{2\pi i}\mu^{1+2s}\int_{\gamma}\mathrm{d}\lambda\lambda^{-s}R(\lambda),
\end{equation}
with resolvent \(R(\lambda)=\frac{\mathrm{d}}{\mathrm{d}\lambda}\ln\Delta(\lambda)\).
The integrand has a branch cut along the negative real axis and poles at the positions of the zeros of \(\Delta(\lambda)\).
The contour \(\gamma\) runs counterclockwise from \(+\infty+i\varepsilon\) to the smallest eigenvalue, crosses the real axis between zero and the
smallest eigenvalue and returns to \(+\infty-i\varepsilon\). Using the residue theorem, one obtains the original definition of the
zeta function (\ref{eq:Zetadef}).

Depending on the behaviour of \(R(\lambda)\) at infinity, for suitable values of \(s\) the contour can now be deformed to lie just above and below 
the branch cut. One gets \cite{Bra1}
\begin{equation}\label{eq:ZetaIntLambda}
 \zeta_D(s)=-\frac{\sin(\pi s)}{\pi}\mu^{1+2s}\int_0^{\infty}\mathrm{d}\lambda\lambda^{-s}R(-\lambda).
\end{equation}
In terms of the momentum variable \(\kappa\) (see (\ref{eq:momentum})) this expression is rewritten as
\begin{equation}\label{eq:ZetaIntKappa}
 \zeta_D(s)=-\frac{\sin(\pi s)}{\pi}\mu^{1+2s}\int_{m}^{\infty}\mathrm{d}\kappa(\kappa^2-m^2)^{-s}\mathcal{R}(\kappa),
\end{equation}
with
\begin{equation}
 \mathcal{R}(\kappa)=\left.2\kappa R(-\lambda)\right|_{\lambda=\kappa^2-m^2}.
\end{equation}
In deriving (\ref{eq:ZetaIntLambda}) we have changed \(\lambda\to -\lambda\), which corresponds to \(\kappa\to i\kappa\). So the correct substitution
of the integration variable in (\ref{eq:ZetaIntLambda}) is \(\kappa^2=\lambda+m^2\) to get (\ref{eq:ZetaIntKappa}).

A pole at \(s=-\frac{1}{2}\) is related to the divergence of the integral (\ref{eq:ZetaIntKappa}) in the upper limit. The divergent parts can be 
isolated by asymptotic expansion of \(\mathcal{R}(\kappa)\) for \(\kappa\to\infty\):
\begin{equation}\label{eq:Asymptotic1}
 \mathcal{R}(\kappa)\to r_0+\frac{r_1}{\kappa^2}+\mathcal{O}(\kappa^{-4}).
\end{equation}
In the case of the sine-Gordon model the first two terms of the expansion are the only divergent contributions to the integral.
Inserting the asymptotic form (\ref{eq:Asymptotic1}) back into (\ref{eq:ZetaIntKappa}) and
making a Laurent expansion for \(s=-\frac{1}{2}+\varepsilon\) around \(\epsilon=0\) the divergence can be made explicit:
\begin{eqnarray}\label{eq:divterms}
 \lim_{s\to-\frac{1}{2}}E_{div}^{(1)}(s)&=&-\frac{r_0m^2}{8\pi}\left[\left.\frac{2}{2s+1}\right|_{s\to-\frac{1}{2}}-1+
2\ln\left(\frac{2\mu}{m}\right)\right],\nonumber\\
 \lim_{s\to-\frac{1}{2}}E_{div}^{(2)}(s)&=&-\frac{r_1}{2\pi}\left[-\left.\frac{1}{2s+1}\right|_{s\to-\frac{1}{2}}+1-
 \ln\left(\frac{2\mu}{m}\right)\right].
\end{eqnarray}
Applying the large mass subtraction condition (\ref{eq:RenormCond}), we have to discard these terms completely 
\begin{equation}
 E_{ren}=\lim_{s\to-\frac{1}{2}}\left[E_{1-loop}(s)-E_{div}^{(1)}(s)-E_{div}^{(2)}(s)\right]
\end{equation}
One can show that these two subtractions are equivalent to the perturbative vacuum and mass renormalization \cite{Pawe}.
The renormalized 1-loop energy contribution is then given by
\begin{equation}
 E_{ren}(k)=\frac{1}{2\pi}\int_m^{\infty}\mathrm{d}\kappa\sqrt{\kappa^2-m^2}\left[\mathcal{R}(\kappa)-r_0-\frac{r_1}{\kappa^2}\right].
\end{equation}
In the following we have to determine \(\mathcal{R}(\kappa)\) and the coefficient \(r_0\) and \(r_1\) for the two boundary conditions separately.

\subsection{Spectral discriminant for \(n=1\) \text{ Lam\'e equation}}
As we will see, the fluctuation equation around the previously presented solutions (\ref{eq:Sinegordonperi}) and (\ref{eq:Sinegordonaperi}) is the
\(n=1\) Lam\'e equation
\begin{equation}\label{eq:Lamestandard}
 -\frac{\mathrm{d}^2f}{\mathrm{d}x^2}+2k^2\mathrm{sn}^2(x,k)f(x)=hf(x)
\end{equation}

For second order differential operators \(-\mathrm{d}_x^2+V(x)\) with periodic potential \(V(x+R)=V(x)\) the discriminant \(\Delta(h)\)
is an entire function of \(h\) and has the general form \cite{Bra1, Bra2, Novi, Kohn, Smir}
\begin{equation}\label{eq:discriminant}
 \Delta(h)=2\cos(Rp(h))\pm 2,
\end{equation}
where the negative and positive signs correspond to periodic and antiperiodic solutions, respectively and 
\(p(h)\) is the quasi-momentum defined by
\begin{equation} 
 f_{h}(x+R)=e^{\pm ip(h)}f_{h}(x).
\end{equation}
The resolvent for e.g. the antiperiodic spectrum is then given by
\begin{equation}
 R(h)=-\tan\left(\frac{R}{2}p(h)\right)p'(h).
\end{equation}
The general solution for (\ref{eq:Lamestandard}) is given by \cite{Whit}
\begin{equation}\label{eq:Lame1Sol}
 f(x)=\frac{H(x+\alpha_1)}{\Theta(x)}e^{-xZ(\alpha_1)},
\end{equation}
(\(H(x), \Theta(x)\) and \(Z(x)\) are the Jacobi eta, theta and zeta function, respectively \cite{Whit, Erde}) provided the additional parameter fulfills the following 
Bethe equation
\begin{equation}
 \mathrm{cn}^2(\alpha_1)\mathrm{ds}^2(\alpha_1)-\mathrm{ns}^2(\alpha_1)=-h.
\end{equation}
The solution is obtained simply by inversion \cite{Bra1}
\begin{equation}\label{eq:BetheSol}
 \alpha_1=\mathrm{sn}^{-1}\left(\sqrt{1-\frac{h-1}{k^2}},k\right).
\end{equation}
The quasi-momentum of the \(n=1\) Lame equation is well known (unlike the case \(n=2\), see \cite{Pawe}) and given by \cite{Bra1}
\begin{equation}\label{eq:QuasiMomentum}
 p(\alpha_1)=iZ(\alpha_1)+\frac{\pi}{2\mathbf{K}},
\end{equation}
which can be obtained from (\ref{eq:Lame1Sol}). Inserting (\ref{eq:BetheSol}) into (\ref{eq:QuasiMomentum}) gives the quasi-momentum \(p(h)\) as
function of \(h\) and the first derivative of the quasi-momentum with respect to \(h\) is given by
\begin{equation}
 p'(h)=\frac{i}{2}\frac{h-\mu_1}{\sqrt{(h_1-h)(h_2-h)(h_3-h)}}
\end{equation}
with
\begin{eqnarray}
 \mu_1&=&k^2+\frac{\mathbf{E}(k)}{\mathbf{K}(k)},\\
 h_1&=&1,\;\;h_2=k^2,\;\;h_3=1+k^2
\end{eqnarray}
Next we need the resolvent \(\mathcal{R}(\kappa)\) (see (\ref{eq:ZetaResolvent})), which has to be considered for the periodic and anti-periodic case 
separately.

\subsection{Periodic}
Expanding the Lagrangian (\ref{eq:Lagrange}) about the periodic solution (\ref{eq:Sinegordonperi}) leads to the following fluctuation 
equation
\begin{equation}\label{eq:PhysFluctPeri}
 \left[-\frac{\mathrm{d}^2}{\mathrm{d}x^2}+2m^2\mathrm{sn}^2\left(\frac{mx}{k},k\right)-m^2\right]\chi(x)=\lambda\chi(x).
\end{equation}
with periodic boundary conditions \(\chi(x+R)=\chi(x)\). This can be brought to the standard form of the \(n=1\) Lame equation 
(\ref{eq:Lamestandard}):
\begin{equation}\label{eq:FluctPeri}
 \left[-\frac{\mathrm{d}^2}{\mathrm{d}\bar x^2}+2k^2\mathrm{sn}^2\left(\bar x,k\right)\right]\chi(\bar x)=h\chi(\bar x),
\end{equation}
with \(\bar x=\frac{m}{k}x\) and
\begin{equation}\label{eq:shiftperi}
 h=k^2\left(1+\frac{\lambda}{m^2}\right)=k^2(2+\frac{\kappa^2}{m^2}).
\end{equation}
In order to apply the results of the previous section we have to recognise the shift (\ref{eq:shiftperi}) for the physical eigenvalues of the periodic 
fluctuations and get as resolvent for the integral representation of the corresponding spectral zeta function  
\begin{equation}\label{eq:resolventkappa1}
 \mathcal{R}(\kappa)=-\kappa R\coth\left(\frac{R}{2}\tilde p(\kappa)\right)\frac{\kappa^2+\mu_1}{\sqrt{(\kappa^2+\kappa_1^2)(\kappa^2+\kappa_2^2)
 (\kappa^2+\kappa_3^2)}}
\end{equation}
 with
\begin{equation}
 \tilde p(\kappa)=\frac{m}{k}\left[Z\left(\mathrm{sn}^{-1}\left(\sqrt{-1+\frac{1}{k^2}+\frac{\kappa^2}{m^2}},k\right),k\right)+\frac{\pi}{2\mathbf{K}}
 \right]
\end{equation}
and
\begin{eqnarray}
 \mu_1=\frac{m^2}{k^2}\left(\frac{\mathbf{E}(k)}{\mathbf{K}(k)}-k^2\right),\\
 \kappa_1^2=\frac{m^2}{k^2}(1-2k^2),\;\;\kappa_2^2=-m^2,\;\;\kappa_3^2=\frac{m^2}{k^2}(1-k^2).
\end{eqnarray}
From (\ref{eq:resolventkappa1}) we can now obtain the coefficients of the asymptotic expansion (\ref{eq:Asymptotic1})
\begin{eqnarray}
 r_0&=&-R=-\frac{2k}{m}\mathbf{K}(k),\\
 r_1&=&-R\left[\mu_1-\frac{1}{2}(\kappa_1^2+\kappa_2^2+\kappa_3^2)\right]=\frac{2m}{k}\left((1-k^2)\mathbf{K}(k)-\mathbf{E}(k)\right).
\end{eqnarray}
Applying the large mass subtraction condition (\ref{eq:RenormCond}), the final renormalized 1-loop energy contribution is then given by
\begin{equation}
 E_{ren}(k)=\frac{1}{2\pi}\int_m^{\infty}\mathrm{d}\kappa\sqrt{\kappa^2-m^2}\left[\mathcal{R}(\kappa)-r_0-\frac{r_1}{\kappa^2}\right].
\end{equation}
\begin{figure}
 \includegraphics[scale=0.8]{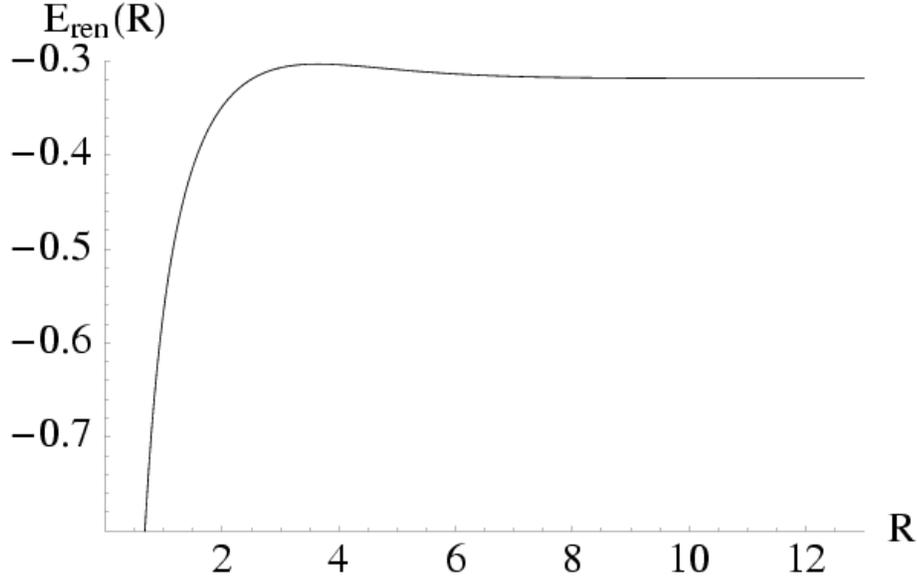}
 \caption{The renormalized 1-loop energy \(E_{ren}(R)\) for the periodic sine-Gordon soliton on \(S^1\) with \(m=1\)}
\end{figure} 
For \(k\to 1\) or \(R\to\infty\) the 1-loop energy approaches the standard result for the Sine-Gordon-Soliton (see Figure 1)
\begin{equation}
 E_{ren}(k)\to -\frac{m}{\pi}.
\end{equation}

\subsection{Anti-periodic}
For anti-periodic boundary condition we have to distinguished between the cases \(R>R_0\) and \(R<R_0\) since only for \(R>R_0\) 
the soliton exists.

\subsubsection{Regularisation for \(R<R_0\)}
For \(R<R_0\) the fluctuation equation for \(\phi=\frac{\pi}{\beta}\) is 
\begin{equation}
 \left[-\frac{\mathrm{d}^2}{\mathrm{d}x^2}-m^2\right]\chi(x)=\lambda\chi(x)
\end{equation}
with anti-periodic spectrum and corresponding spectral discriminant
\begin{equation}\label{eq:eigenval}
 \lambda_n=\left(\frac{(2n+1)\pi}{R}\right)^2-m^2\;\;\;\;\Leftrightarrow\;\;\;\Delta(\lambda)=\cos^2\left(\frac{R}{2}\sqrt{\lambda+m^2}\right).
\end{equation}
The integral representation of spectral zeta function is given by (\ref{eq:ZetaIntLambda}) with
\begin{equation} 
 R(-\lambda)=-\frac{R}{2}\frac{\tanh\left(\frac{R}{2}\sqrt{\lambda-m^2}\right)}{\sqrt{\lambda-m^2}}.
\end{equation}
In this expression we have already deformed the integration contour from the poles on the positive real axis to the
branch cut along the negative real axis. This is valid for \(\frac{1}{2}<\mathrm{Re}(s)<1\) and \(mR<\pi\). The restriction \(mR<\pi\) is necessary since
for fixed radius \(R\) the first eigenvalues (\ref{eq:eigenval}) become negative when \(mR\) becomes larger than \(\pi\) and
the corresponding poles of \(R(\lambda)\) move into the branch cut. This makes the integral representation invalid.
Using the momentum-like parameter (\ref{eq:momentum}) we get
\begin{equation}\label{eq:intrepzeta}
 \zeta_D(s)=-\mu^{1+2s}\frac{\sin(\pi s)}{\pi}\int_{m}^{\infty}\mathrm{d}\kappa(\kappa^2-m^2)^{-s}\mathcal{R}(\kappa),
\end{equation} 
with 
\begin{equation}
 \mathcal{R}(\kappa)=-\frac{R\kappa\tanh\left(\frac{R}{2}\sqrt{\kappa^2-2m^2}\right)}{\sqrt{\kappa^2-2m^2}}.
\end{equation}
The renormalization condition (\ref{eq:RenormCond}) cannot be applied in this case, since for fixed \(R\) we cannot take \(m\to\infty\).
Instead we have first to renormalized the 1-loop energy in the \(R>R_0\) region and then to impose the condition that the renormalized 1-loop
energy for \(R<R_0\) and \(R>R_0\) have to be continuous at \(R=R_0\).

\subsubsection{Regularization and renormalization for \(R>R_0\)}
For \(R>R_0\) the fluctuation equation around the anti-periodic configuration (\ref{eq:Sinegordonaperi}) is given by
\begin{equation}
 \left[-\frac{\mathrm{d}^2}{\mathrm{d}x^2}+2m^2k^2\mathrm{sn}^2(mx)-m^2\right]\chi(x)=\lambda\chi(x).
\end{equation}
This can be brought to the standard form of the \(n=1\) Lame equation (\ref{eq:Lamestandard})
\begin{equation}
 \left[-\frac{\mathrm{d}^2}{\mathrm{d}\bar x^2}+2k^2\mathrm{sn}^2(\bar x)\right]\chi(\bar x)=h\chi(\bar x),
\end{equation}
with \(\bar x=mx\) and
\begin{equation}\label{eq:shiftaperi}
 h=\frac{\lambda}{m^2}+1=\frac{\kappa^2}{m^2}+2.
\end{equation}
In order to apply the results of the previous section we have to recognise the shift (\ref{eq:shiftaperi}) for the physical eigenvalues of the 
anti-periodic fluctuations and get as resolvent for the integral representation of the corresponding spectral zeta function
\begin{equation}\label{eq:resolventkappa2}
 \mathcal{R}(\kappa)=-\kappa R\tanh\left(\frac{R}{2}\tilde p(\kappa)\right)\frac{\kappa^2+\mu_1}{\sqrt{(\kappa^2+\kappa_1^2)(\kappa^2+\kappa_2^2),
 (\kappa^2+\kappa_3^2)}},
\end{equation}
with
\begin{equation}
 \tilde p(\kappa)=m\left[Z\left(\mathrm{sn}^{-1}\left(\sqrt{1-\frac{1}{k^2}+\frac{\kappa^2}{m^2k^2}},k\right),k\right)+\frac{\pi}{2\mathbf{K}}\right],
\end{equation}
and 
\begin{eqnarray}
 \mu_1=m^2\left(\frac{\mathbf{E}(k)}{\mathbf{K}(k)}+k^2-2\right),\\
 \kappa_1^2=-m^2,\qquad\kappa_2^2=m^2(k^2-2),\qquad\kappa_3^2=m^2(k^2-1).
\end{eqnarray}
The coefficients in the asymptotic expansion (\ref{eq:Asymptotic1}) of (\ref{eq:resolventkappa2}) are
\begin{eqnarray}
 r_0&=&-R=-\frac{2}{m}\mathbf{K}(k),\\
 r_1&=&-2m\mathbf{E}(k).
\end{eqnarray}

\begin{figure}\label{fig:aperi}
 \includegraphics[scale=0.8]{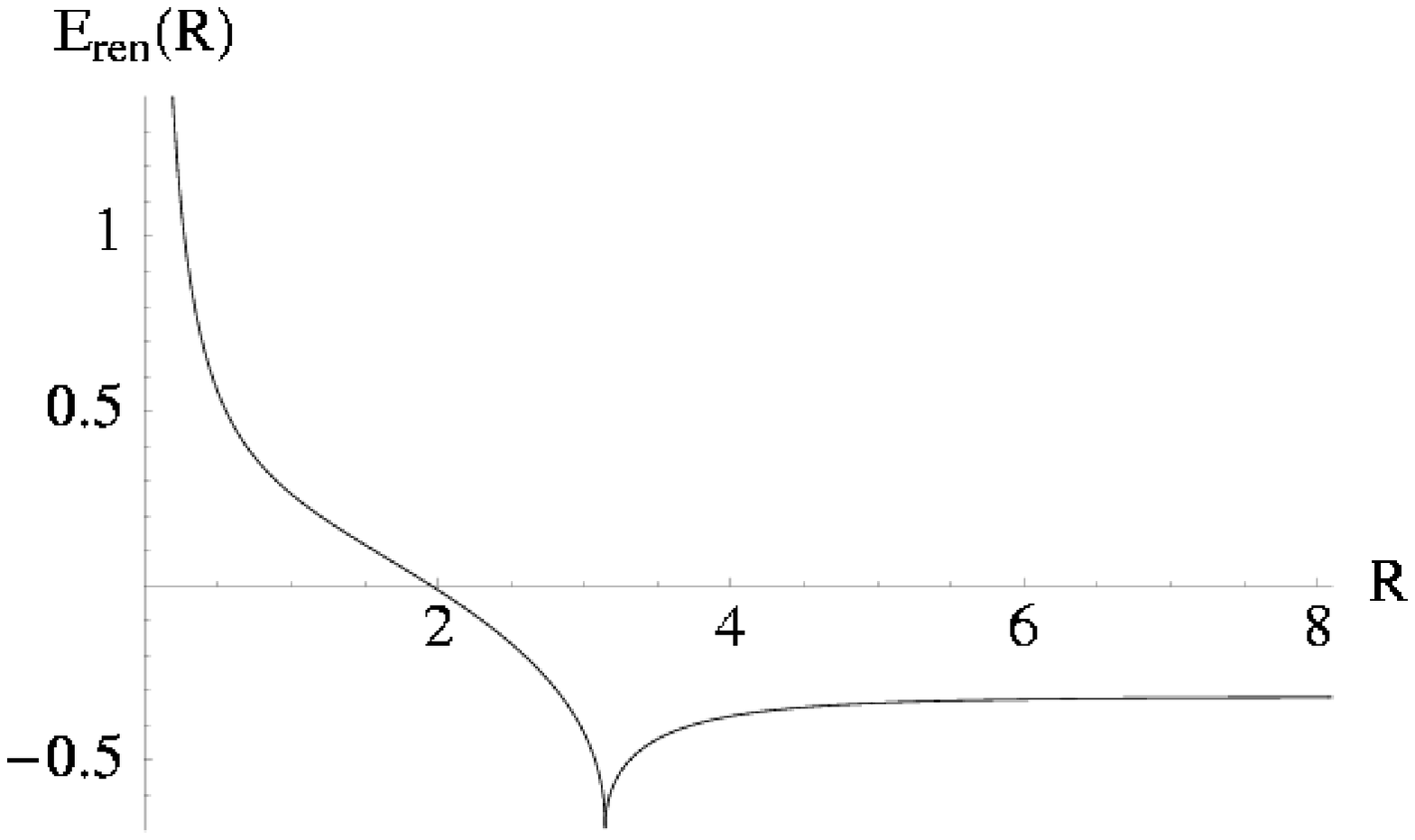}
 \caption{The renormalized 1-loop energy \(E_{ren}(R)\) for the anti-periodic sine-Gordon soliton on \(S^1\) with \(m=1\)}
\end{figure} 

Applying the large mass subtraction scheme (\ref{eq:RenormCond}) the renormalized 1-loop energy contribution is given by
\begin{equation}\label{eq:aperi1}
 E_{ren}(k)=\frac{1}{2\pi}\int_m^{\infty}\mathrm{d}\kappa\sqrt{\kappa^2-m^2}\left[\mathcal{R}(\kappa^2)-r_0-\frac{r_1}{\kappa^2}\right].
\end{equation}
As in the case for the periodic soliton, for \(k\to 1\) (or \(R\to\infty\)) the 1-loop energy approaches the standard result for the 
sine-Gordon soliton (see Figure 2)
\begin{equation}
 E_{ren}(k)\to -\frac{m}{\pi}.
\end{equation}

\subsubsection{Renormalization for \(R<R_0\)}
We have seen that the large mass renormalization condition (\ref{eq:RenormCond}) cannot applied to (\ref{eq:intrepzeta}) for \(R<R_0\), 
but now we have a renormalized result for the energy for \(R>R_0\) and a natural renormalization condition for \(R<R_0\) is that the renormalized 
energy for \(R<R_0\) has to match at \(R=R_0\) the renormalized energy for \(R>R_0\):
\begin{equation}\label{eq:newrenormcond}
 E_{ren,R<R_0}(R)\to E_{ren,R>R_0}(R_0) \text{ for } R\to R_0
\end{equation}
The terms which have to be subtracted for \(R<R_0\) from (\ref{eq:intrepzeta}) can then be identified by the renormalization condition (\ref{eq:newrenormcond}) as
\begin{eqnarray}
 \lim_{s\to -\frac{1}{2}}E_{div}^{(1)}(s)&=&\frac{Rm^2}{8\pi}\left[\left.\frac{2}{2s+1}\right|_{s\to-1/2}-1+2\ln\left(\frac{2\mu}{m}\right)\right],
 \nonumber\\
 \lim_{s\to -\frac{1}{2}}E_{div}^{(2)}(s)&=&\frac{Rm^2}{2\pi}\left[-\left.\frac{1}{2s+1}\right|_{s\to-1/2}+1-\ln\left(\frac{2\mu}{m}\right)\right].
\end{eqnarray}

In the sector \(R<R_0\) we get therefore the renormalized 1-loop contribution (see Figure 2)
\begin{equation}\label{eq:aperi2}
 E_{ren}(R)=-\frac{R}{2\pi}\int_m^{\infty}\mathrm{d}\kappa\sqrt{\kappa^2-m^2}\left[
 \frac{\kappa\tanh\left(\frac{R}{2}\sqrt{\kappa^2-2m^2}\right)}{\sqrt{\kappa^2-2m^2}}-1-\frac{m^2}{\kappa^2}\right].
\end{equation}

\section{Discussion}

\begin{figure}\label{fig:aperiphys}
 \includegraphics[scale=0.8]{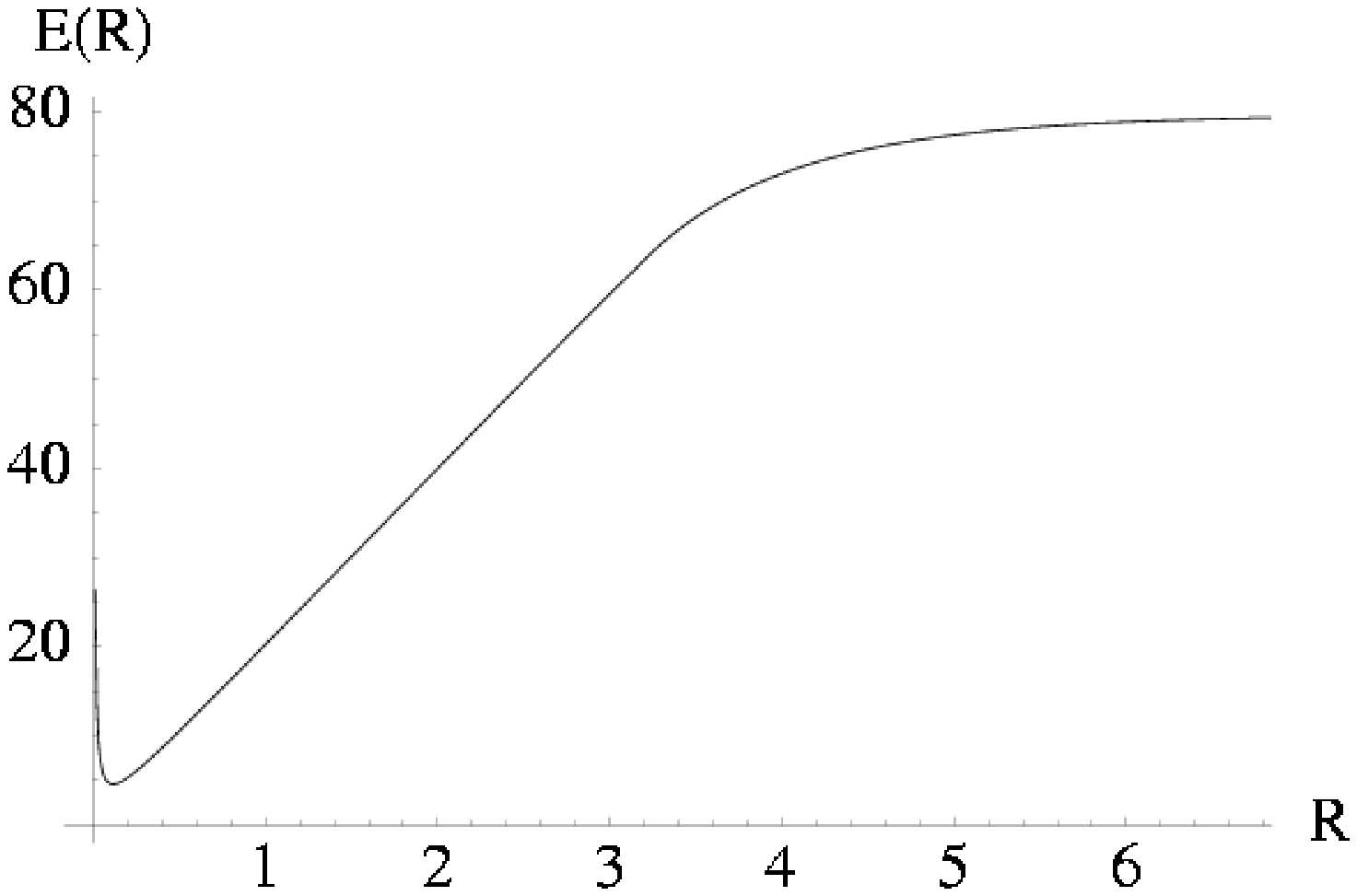}
 \caption{The physical energy \(E(R)\) for the anti-periodic sine-Gordon soliton on \(S^1\) with \(m=1\) and \(\beta^2=0.1\)}
\end{figure}
\begin{figure}\label{fig:aperiphys2}
 \includegraphics[scale=0.8]{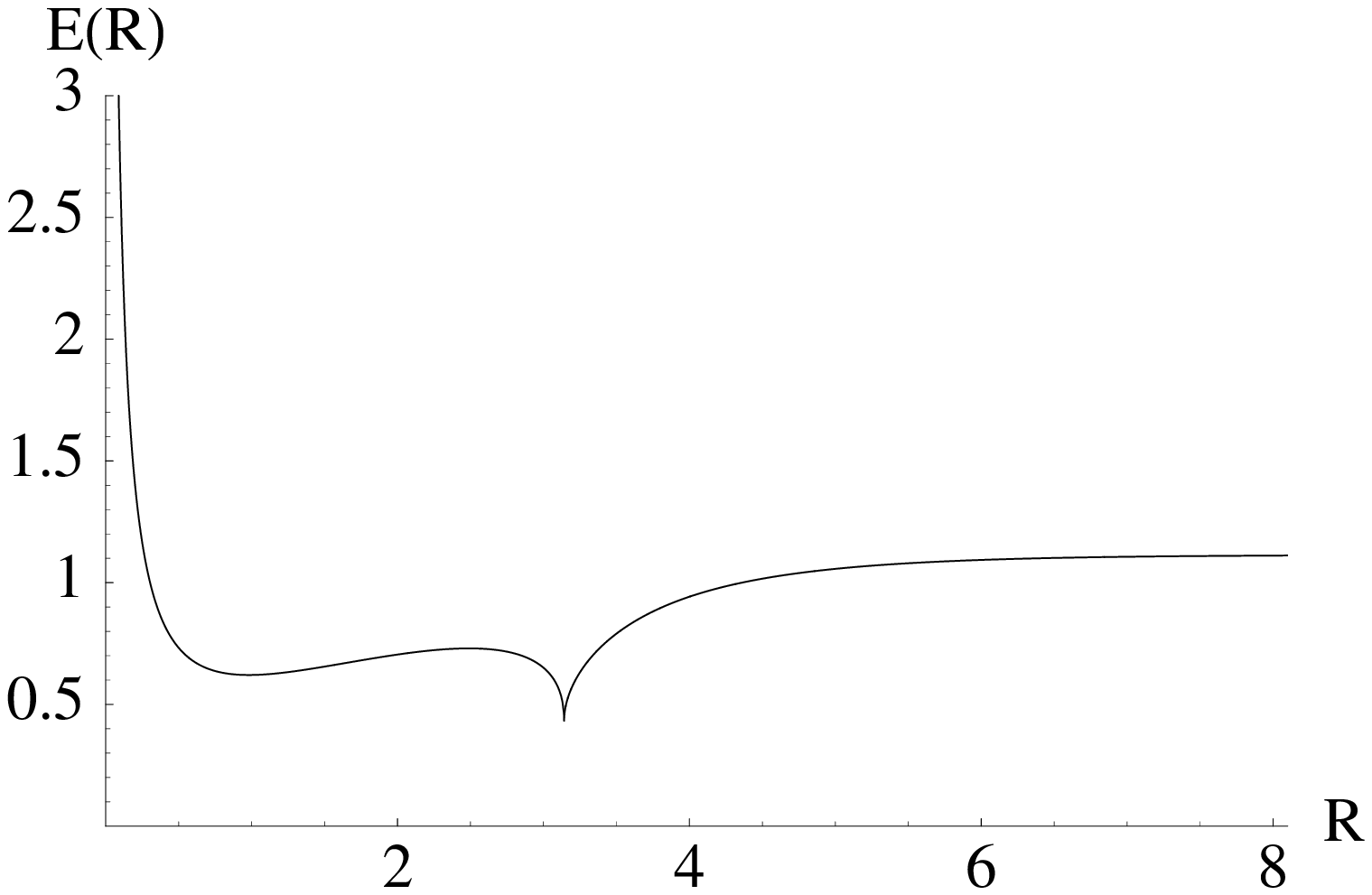}
 \caption{The physical energy \(E(R)\) for the anti-periodic sine-Gordon soliton on \(S^1\) with \(m=1\) and \(\beta^2=5.585\)}
\end{figure}

\begin{figure}\label{fig:periphys}
 \includegraphics[scale=0.8]{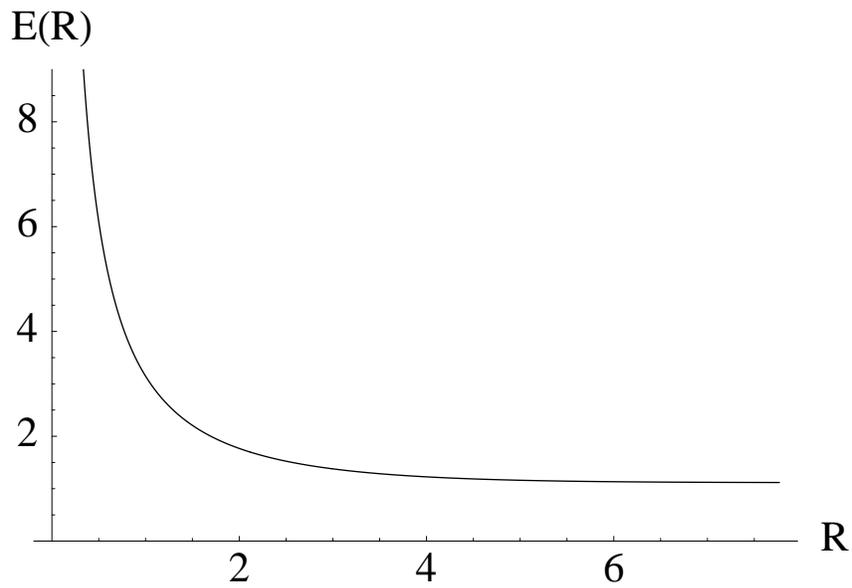}
 \caption{The physical energy \(E(R)\) for the periodic sine-Gordon soliton on \(S^1\) with \(m=1\) and \(\beta^2=5.585\)}
\end{figure}
Our semiclassical results are at first valid as long as \(\beta^2\ll 1\), which is our dimensionless expansion parameter.
In Figure 3 and 4 the physical energy 
\begin{equation}\label{eq:physicalenergy}
E(R)=E_{cl}(R)+E_{ren}(R)
\end{equation}
is plotted for the anti-periodic case for \(\beta^2=0.1\) and \(\beta^2=5.585\), respectively. The critical radius above the soliton can exist lies at \(R_0=\pi\).
One can see that also a minimum in the physical energy appears for 
\begin{equation}
 R_{min}\approx\frac{1}{2m}\beta.
\end{equation}
Since \(R_{min}<R_0\) the minimum appears in the homogeneous phase \(\phi_0=\pi/\beta\). 
The significant cusp at \(R=R_0\) seen in Figure 4 we interpret as an indication of the breakdown of the semiclassical approximation, which is
expected for a value of \(\beta^2=5.585\), and higher loop effects have to take into account, at least around \(R_0\). 
This is qualitatively the same behaviour as for the \(\phi^4\) kink on \(S^1\) with anti-periodic b.c. \cite{Pawe}.

For periodic boundary conditions a soliton solution exist for all values of \(R\). The physical energy \(E(R)\) is plotted in
Figure 5 for \(\beta^2=5.585\). By setting \(l=E_{QS}R\) with the quantum soliton mass \(E_{QS}=8m/\beta^2-m/\pi\) we can compare (see Table I) our
semiclassical result \(E(l)/E_{QS}\) with the numerically determined exact values \cite{Feve} using the 
integrability of the sine-Gordon model \cite{Fev2}. We find an astonishing agreement far outside the semiclassical regime for 
\(\beta^2=5.585\). The maximal relative deviation between the exact numerical NLIE (non linear integral equations) result and the 
semiclassical value around \(l=4\) can be understood, if one considers the radiative corrections, which travel around the compact dimension of 
circumference \(R\). These are additional loop contributions, which are not present when the soliton lives on an infinite line. Their contribution 
is maximal when \(1/R\) is of the same order as the mass \(m\) of the fluctuating particles \cite{Lusc}. Let us take for concreteness the critical 
radius \(R_0=\pi/m\). Then the expected value for \(l\) where this contributions become maximal for \(\beta^2=5.585\) lies at 
\(l_0=E_{QS}R_0=8\pi/\beta^2-1\approx 3.5\), which is in good agreement with the numerical results in Table I.

\begin{table}
\caption{\label{tab:list}Numerical comparison of \(E(l)/E_{QS}\) obtained semiclassical and by the NLIE method \cite{Feve} for \(\beta^2=5.585\) and
periodic boundary conditions}

\begin{tabular}{@{}c|l|l|l||l|l|l}

 \(l\) & \(E_{cl}\) & \(E_{ren}\) & \(\frac{|E_{ren}|}{E_{cl}}\) & \(E/E_{QS}\) semicl. & \(E/E_{QS}\) with NLIE & relative deviation\\ \hline\hline
 0.5 & 7.95543 & -1.18475 & 0.149  & 6.07727 & 6.080571 & 0.0005\\
 1 & 4.09746 & -0.62027 & 0.1514 & 3.12108 & 3.126706 & 0.0018\\
 1.5 & 2.86336 & -0.44559 & 0.1556 & 2.17017 & 2.177411 & 0.0033\\
 2 & 2.28367 & -0.36882 & 0.1615 & 1.71874 & 1.727224 & 0.0049\\
 2.5 & 1.96405 & -0.33127 & 0.1687 & 1.46556 & 1.475004 & 0.0064\\
 3 & 1.77267 &-0.31298 & 0.1766 & 1.31020 & 1.320353 & 0.0077\\
 4 & 1.57577 & -0.30291 & 0.1922 & 1.14250 & 1.153188 & 0.0093\\
 5 & 1.49304 & -0.30588 & 0.20487 & 1.06557 & 1.075376 & 0.0091\\
\end{tabular}

\end{table}

The quantum corrections to the physical energy of the soliton cannot be called small in any sense for \(\beta^2=5.585\), since as one can also see 
from Table I already the 1-loop corrections \(E_{ren}\) have an effect up to 20 per cent compared to the classical part \(E_{cl}\). Nevertheless the semiclassical result is a good approximation to the exact
values since higher loop effects only accumulate into a contribution \(\delta E_{h.l.}=E_{\text{by NLIE}}-E_{\text{by s.cl.}}\) of at most 0.7 and 4 per cent 
at \(l=4\) compared to \(E_{cl}\) and \(E_{ren}\), respectively and even decrease when going \(l\to 0\).

If we now make the assumption that the ratio \(\delta E_{h.l}/E_{ren}\) is nearly the same for the different boundary conditions, we can
make a conjecture about the magnitude of the higher loop contributions in the anti-periodic case of (\ref{eq:physicalenergy}). For \(l=1\), where the 
estimated minimum lies, we get for \(\delta E_{h.l.}/(E_{cl}+E_{ren})\) a value about 0.5 per cent, which means that the observed minimum has a chance to be 
physically valid.
\begin{table}
\caption{\label{tab:list}Numerical comparison of \(E_{ren}(R)\) for the periodic soliton and the Casimir energy of a massless free field for small \(R\)}
\begin{tabular}{@{}c|l|l}

 \(R\) & \(E_{ren}(R)\) & \(-\pi/6R\)\\ \hline\hline
 0.000628 & -833.333 & -833.333\\
 0.001257 & -416.667 & -416.667\\
 0.001885 & -277.778 & -277.778\\
 0.003142 & -166.667 & -166.667\\
 0.004398 & -119.048 & -119.048\\
 0.005027 & -104.167 & -104.167\\
 0.006283 & -83.3335 & -83.3332\\
 0.007540 & -69.4446 & -69.4443\\
\end{tabular}

\end{table}

Finally, we mention that for \(R\to 0\) the renormalized 1-loop contribution \(E_{ren}(R)\) for periodic b.c. approaches the Casimir energy of a free massless field:
\begin{equation}
 E_{ren}(R)\to -\frac{\pi}{6R},\qquad R\to 0,
\end{equation}
since the fluctuation spectrum following from (\ref{eq:PhysFluctPeri}) becomes in leading order for \(R\to 0\)
\begin{equation} 
 \lambda_n\to\frac{(2\pi n)^2}{R^2},
\end{equation}
which is the spectrum of a free massless scalar field. A Numerical calculation shows this behaviour (see Table II).
Therefore the physical energy approaches in this limit
\begin{equation}
 E(R)\to\frac{2\pi}{R}\left(\frac{\pi}{\beta^2}-\frac{1}{12}\right),\qquad R\to 0.
\end{equation}
without significant higher loop corrections.

\section{Conclusion}
In this letter we have applied the techniques of \cite{Pawe} in order to obtain analytic results for the 1-loop quantum mass correction of the sine-Gordon soliton on 
\(S^1\) with (anti-) periodic boundary conditions. Since the sine-Gordon model is integrable we were able to compare the semiclassical results
with exact numerical ones in the case of periodic boundary condition. We found in this case that the semiclassical approximation gives very good
results even outside the expected region of validity of the semiclassical method.

In the case of anti-periodic boundary conditions a radius of minimal energy was semiclassical obtain, since the classical and 1-loop contributions
are of same magnitude at this point. By learning from the periodic case we have conjectured that the higher loop contributions at this point will
nevertheless be insignificant and the obtained minimum therefore physically valid.

It would be interesting, if the NLIE method \cite{Fev2, Feve} can also be applied to anti-periodic boundary conditions, in order to test the semiclassical 
predictions of this letter.

\end{document}